\documentclass[12pt]{article}
\usepackage{graphicx}

\def\Re{{\cal R \mskip-4mu \lower.1ex \hbox{\it e}\,}}
\def\Im{{\cal I \mskip-5mu \lower.1ex \hbox{\it m}\,}}
\def\ie{{\it i.e.}}
\def\eg{{\it e.g.}}

\def\sub#1{_{\lower.25ex\hbox{$\scriptstyle#1$}}}
\def\tev{\,{\ifmmode\mathrm {TeV}\else TeV\fi}}
\def\gev{\,{\ifmmode\mathrm {GeV}\else GeV\fi}}
\def\mev{\,{\ifmmode\mathrm {MeV}\else MeV\fi}}
\def\mpl{\ifmmode M_{pl}\else $M_{pl}$\fi}
\def\to{\rightarrow}

\def\subw{_{\rm w}}
\def\mh{\ifmmode m\sbl H \else $m\sbl H$\fi}
\def\mch{\ifmmode m_{H^\pm} \else $m_{H^\pm}$\fi}
\def\mt{\ifmmode m_t\else $m_t$\fi}
\def\mc{\ifmmode m_c\else $m_c$\fi}
\def\mz{\ifmmode M_Z\else $M_Z$\fi}
\def\mw{\ifmmode M_W\else $M_W$\fi}
\def\mws{\ifmmode M_W^2 \else $M_W^2$\fi}
\def\mhs{\ifmmode m_H^2 \else $m_H^2$\fi}   
\def\mzs{\ifmmode M_Z^2 \else $M_Z^2$\fi}
\def\mts{\ifmmode m_t^2 \else $m_t^2$\fi}
\def\mcs{\ifmmode m_c^2 \else $m_c^2$\fi}
\def\mchs{\ifmmode m_{H^\pm}^2 \else $m_{H^\pm}^2$\fi}
\def\ztwo{\ifmmode Z_2\else $Z_2$\fi}
\def\zone{\ifmmode Z_1\else $Z_1$\fi}
\def\mtwo{\ifmmode M_2\else $M_2$\fi}
\def\mone{\ifmmode M_1\else $M_1$\fi}
\def\tb{\ifmmode \tan\beta \else $\tan\beta$\fi}
\def\xw{\ifmmode x\subw\else $x\subw$\fi}
\def\ch{\ifmmode H^\pm \else $H^\pm$\fi}
\def\lum{\ifmmode {\cal L}\else ${\cal L}$\fi}
\def\inpb{\,{\ifmmode {\mathrm {pb}}^{-1}\else ${\mathrm {pb}}^{-1}$\fi}}
\def\infb{\,{\ifmmode {\mathrm {fb}}^{-1}\else ${\mathrm {fb}}^{-1}$\fi}}
\def\epem{\ifmmode e^+e^-\else $e^+e^-$\fi}
\def\ppb{\ifmmode \bar pp\else $\bar pp$\fi}
\def\bsg{\ifmmode B\to X_s\gamma\else $B\to X_s\gamma$\fi}
\def\bsll{\ifmmode B\to X_s\ell^+\ell^-\else $B\to X_s\ell^+\ell^-$\fi}
\def\bstt{\ifmmode B\to X_s\tau^+\tau^-\else $B\to X_s\tau^+\tau^-$\fi}
\def\lamt{\ifmmode \tilde\lambda\else $\tilde\lambda$\fi}
\def\shat{\ifmmode \hat s\else $\hat s$\fi}
\def\that{\ifmmode \hat t\else $\hat t$\fi}
\def\uhat{\ifmmode \hat u\else $\hat u$\fi}

\newskip\zatskip \zatskip=0pt plus0pt minus0pt
\def\matth{\mathsurround=0pt}
\def\lsim{\mathrel{\mathpalette\atversim<}}

\def\atversim#1#2{\lower0.7ex\vbox{\baselineskip\zatskip\lineskip\zatskip
  \lineskiplimit 0pt\ialign{$\matth#1\hfil##\hfil$\crcr#2\crcr\sim\crcr}}}

\renewcommand{\thefootnote}{\fnsymbol{footnote}}

\begin{document} \begin{titlepage} 
\rightline{\vbox{\halign{&#\hfil\cr
&SLAC-PUB-9190\cr
&May 2002\cr}}}
\begin{center}

{\Large\bf Radion Couplings to Bulk Fields in the Randall-Sundrum Model}
\footnote{Work supported by the Department of 
Energy, Contract DE-AC03-76SF00515}
\medskip

\normalsize 
{\bf \large Thomas G. Rizzo}
\vskip .3cm
Stanford Linear Accelerator Center \\
Stanford University \\
Stanford CA 94309, USA\\
\vskip .2cm

\end{center}

\begin{abstract} 
The radion may be the lightest new state present in the Randall-Sundrum(RS) 
model. We examine the couplings of the radion to the Standard Model(SM) fields 
in the scenario where they propagate in the bulk and expand into Kaluza-Klein 
towers. These couplings are then contrasted with those of the more familiar 
case where the SM fields are confined to the TeV brane. We find that the 
couplings of the radion to both $gg$ and $\gamma\gamma$ can be significantly 
different in these two cases. Implications for radion collider phenomenology 
are discussed. 
\end{abstract} 



\renewcommand{\thefootnote}{\arabic{footnote}} \end{titlepage}


\section{Introduction}

Theories with extra dimensions provide a new way to attack the hierarchy 
problem. An exciting feature of these scenarios is that they lead to concrete 
and quite distinctive phenomenological tests. If such theories truly describe 
the source of the observed hierarchy, then their direct signatures should 
appear in future collider experiments that probe the TeV scale. In the 
specific model of Randall and Sundrum (RS)\cite{rs}, the observed hierarchy is created by
an exponential warp factor which arises from a 5-dimensional 
non-factorizable geometry. The collider signatures for this model have been 
studied in detail in {\cite {dhr}}. 

The RS setup consists of a 5-dimensional non-factorizable
geometry based on a slice of AdS$_5$ space with length $\pi r_c$, where $r_c$
denotes the compactification radius.  Two 3-branes,
with equal and opposite tensions, rigidly reside at the $S_1/Z_2$ orbifold
fixed points at the boundaries of the AdS$_5$ slice, taken to be 
$y=r_c\phi=0,r_c\pi$.  The 5-dimensional Einstein's
equations permit a solution which preserves 4-dimensional Poincar\' e
invariance with the metric
\begin{equation}
ds^2=e^{-2\sigma(\phi)}\eta_{\mu\nu}dx^\mu dx^\nu - r_c^2d\phi^2\,,
\end{equation}
where the Greek indices extend over ordinary 4-d spacetime and the warp factor 
is given by 
$\sigma(\phi)=kr_c|\phi|$.  Here $k$ is the AdS$_5$ curvature scale which is of
order the Planck mass and is determined by the bulk cosmological constant
$\Lambda=-24M_5^3k^2$, where $M_5$ is the 5-dimensional Planck mass that 
appears in the RS action. 
The 5-d curvature scalar is then given by $R_5=-20k^2$; the requirement that 
quantum corrections be small, $|R_5| < M_5^2$, implies $k/\mpl \lsim 0.1$. 
Examination of the action in the 4-d effective theory yields the relation
$\mpl^2={M_5^3\over k}(1-e^{-2kr_c\pi})\simeq {M_5^3\over k}$
for the reduced 4-d Planck mass.  
The scale of physical phenomena as realized by the 4-d flat metric 
transverse to the 5th dimension $y=r_c\phi$ 
is specified by the exponential warp factor.
TeV scales can naturally be attained on the 3-brane at $\phi=\pi$ if
gravity is localized on the Planck brane at $\phi=0$ and $kr_c\simeq 11-12$. 
(We take $kr_c=11.27$ in our numerical analysis below.) 
The scale of physical processes on this SM or TeV-brane is then 
$\Lambda_\pi\equiv\mpl e^{-kr_c\pi}$.
The observed hierarchy is thus generated by a geometrical exponential
factor and no other additional large hierarchies appear in the model.   

That the quantity $kr_c$ can be stabilized and the above 
range of values can be made natural has 
been demonstrated by a number of authors{\cite {gw}} and leads directly to 
the existence of a massive radion($r$), which corresponds to a 
quantum excitation of 
the brane separation. In the original RS scenario, SM matter was confined 
to the TeV brane. In this case it can be 
shown that the radion couples to the trace of the 
stress-energy tensor of the SM wall fields 
with a strength $\Lambda_r=\sqrt 3 \Lambda_\pi$ which is of 
order the TeV scale, \ie, ${\cal L}_{eff}=-r~T^\mu_\mu /\Lambda_r$. 
This leads to gauge and matter couplings for the radion that 
are qualitatively similar to those of the SM 
Higgs boson. The radion mass ($m_r$), which arises dynamically 
during the stabilization procedure, is expected to be significantly 
below the scale $\Lambda_r$ implying that the radion may be 
the lightest new field present in the RS model. One may expect on general 
grounds that this mass should lie in the range of  
a few $\times 10$ GeV $\leq m_r \leq \Lambda_r$. The phenomenology of the RS 
radion coupled to SM wall fields on the TeV brane has been examined by a 
number of authors{\cite {big}} and in particular has been recently reviewed 
by Kribs{\cite {Kribs}} to which the interested reader should refer for 
details.

For model building reasons one may consider placing some if not all of the 
SM gauge and fermion 
fields in the RS bulk; this possibility has been systematically examined 
by a number of authors{\cite {dhr,big2}} in the non-recoil limit. While it 
has been shown that it is possible to consistently place 
SM gauge bosons and fermions in the bulk, for phenomenological reasons 
the Higgs fields 
which induce $SU(2)_L\times U(1)_Y$ breaking must remain on the TeV brane. In 
such a scenario one may wonder how the radion couplings with the zero-modes 
of the  
bulk fields, which we now identify with the observed SM particles, may be 
modified from those which occur when the SM fields are on the TeV brane. 
The purpose of this paper is to examine this issue in detail and elucidate 
the phenomenology of radions coupling to the bulk SM fields. Since the 
radion is expected to be the lightest new particle in the RS model, its 
couplings to the zero-modes of these fields will be the first probe of the 
detailed structure of the model. In order to isolate 
which new effects are due to `bulk versus brane' couplings, 
we will assume that all mixing 
among the Kaluza-Klein(KK) levels of bulk states is sufficiently small as to be 
negligible.

\section{Bulk Couplings}

When the SM fields are constrained to lie on the TeV brane their couplings to 
the radion only arise through spontaneous symmetry 
breaking(SSB), \ie, through the 
induced masses of the fermions and gauge bosons. In the absence of such terms 
the trace of the 4-d stress-energy tensor would receive zero contribution from 
these sectors. These Higgs-induced terms can be generically written as
\begin{equation}
S_{wall}=\int d^4x ~{{-r}\over {\Lambda_r}}[m_f\bar ff-M_V^2 V_\mu V^\mu]\,,
\end{equation}
where $f$ is any SM fermion and $V=W,Z$. In contrast to this situation, 
when the SM gauge and fermion fields are placed in the bulk, radion couplings 
can arise from a number of sources which we will investigate below. 
The terms of the stress-energy tensor to which the radion 
eventually couples now receive contributions from both bulk and 
brane terms in the full action. However, we note that with the Higgs field 
remaining 
on the TeV brane the gauge and fermion mass terms arising from spontaneous 
symmetry breaking remain wall terms in the action as they are above. In the 
case where the SM fields are in the bulk, SSB must still induce masses for the 
zero modes of these fields which coincide with their SM values. 
This means that $S_{wall}$ will be present in either case and its existence 
is not sensitive to the placement of the fermion and and gauge fields. 
Similarly, from these arguments it is obvious that the radion coupling to 
pairs of Higgs bosons will be left unaltered by the considerations below 
since the Higgs remains on the TeV brane. With this in mind we now turn to the 
{\it pure} bulk terms and ignore those arising from the wall except when they 
are added to complete the final coupling expressions. In effect, this implies 
that we can examine the bulk terms in the absence of any SSB contributions. 

In order to formulate the couplings of the radion to the bulk SM fields we 
must obtain expressions for the dynamical components of the metric tensor 
for the 
full RS model in 5-dimensions in terms of physical fields. This has recently 
been studied in detail in {\cite {roch}} to which we refer the reader for 
details. Removing the pure graviton tower 
excitation pieces from the dynamical parts of the metric yields the 
radion contributions. To linear order in the fields, the $\mu\nu$-components 
of the relevant terms in the metric are found to be  
\begin{equation}
g_{\mu\nu}^r=-e^{2kr_ct}\Big[{{(\pi \epsilon^2+t)\epsilon^{-1}} 
\over {\sqrt 6\pi \mpl}}
\eta_{\mu\nu}~r+{{\tilde S(t)\epsilon^{-1}}\over {\sqrt 24 \pi kr_c\mpl}}
\partial_\mu \partial_\nu r \Big]\,,
\end{equation}
where $\epsilon=e^{-kr_c\pi}$, $t=|\phi|$ and $r$ is the $\phi$-independent 
radion field. The function $\tilde S(t)$ is defined by{\cite {roch}} 
\begin{equation}
\tilde S(t)=\sum_{n=1} \kappa_n [\chi^{(n)}(t)-\chi^{(n)}(\pi)]\,,
\end{equation}
where the $\chi^{(n)}(t)$ are the familiar 
wave functions for the RS graviton tower KK excitations{\cite {dhr}}:
\begin{equation}
\chi^{(n)}(t)\simeq (\sqrt{kr_c} \epsilon^{-1})z^2 J_2(x_n z)/J_2(x_n)\,,
\end{equation}
where $J_2$ is a Bessel function, 
$z=e^{kr_c(t-\pi)}$, with $x_n$ being the roots of $J_1$ and the 
corresponding $Y_2$ terms 
have been dropped for simplicity of notation. The coefficients 
$\kappa_n$ are then given by{\cite {roch}}
\begin{equation}
\kappa_n=2I_n/m_n^2={{-8kr_c}\over {(kx_n\epsilon)^2}}\int_0^\pi d\phi 
~e^{-2kr_c\phi}~\phi \chi^{(n)}(\phi)\,,
\end{equation}
where $m_n=kx_n\epsilon$ are the graviton KK excitation masses. 
Note that we have made use of the gauge freedom discussed in {\cite {roch}} 
so that $\tilde S(t)$, and hence the term proportional to 
$\sim \partial_\mu \partial_\nu r$, 
vanishes on the TeV brane. In this limit the authors of Ref. {\cite {roch}} 
have shown that the `standard' radion interactions are recovered 
when all SM matter lies on the TeV brane. 
Correspondingly, the radion-dependent parts of the $\mu 5$- and 
$55$-components of the metric are given to linear order 
by the much simpler expressions:  
\begin{eqnarray}
g_{55}^r &=& {{\epsilon^{-1}}\over {\sqrt 6 \pi kr_c \mpl}}r\,, \nonumber \\
g_{\mu 5}^r &=& \phi {{\epsilon^{-1}}\over {\sqrt {24} \pi kr_c \mpl}}
\partial_\mu r\,,
\end{eqnarray}
What is interesting about the term arising from the $\mu 5$-components of the 
metric is that it is odd under the 
$Z_2$ orbifolding while all zero-mode bulk fields are $Z_2$ even. As we will 
see below this implies that any radion couplings induced by these terms must 
vanish unless accompanied by derivatives of the compactified coordinate. It 
is clear from the above that the radion couplings to bulk SM fields can arise 
from terms in the stress-energy tensor of four different types:  
$T_\mu^\mu$, as above, 
$T_{\mu\nu}$ contracted with two powers of the radion momenta, 
$T_{55}$, and lastly $T_{\mu 5}$. With the normalization above the radion 
couplings to matter that we are interested in 
are just the product of the $g^r_{AB}$ contracted with 
the corresponding terms in the stress energy tensor produced by the SM KK zero 
modes. Let us now examine each of these terms in 
turn for both massless gauge fields and fermions in the bulk.

\subsection{Bulk Gauge Fields}

Here we are interested in the decay to and the couplings of on-shell radions to 
SM gauge bosons. We recall that in the RS model all gauge fields begin as 
massless in 5-d with spontaneous symmetry breaking taking place on the TeV 
brane. The latter contributions to the radion couplings have already been 
discussed above so here we concentrate solely on the new 5-d pieces ignoring 
the contributions from SSB.  
Suppressing possible gauge group indices the general form of the 5-d 
stress-energy tensor for on-shell (or unitary gauge) {\it massless} bulk 
gauge fields is given by{\cite {feyn}}
\begin{equation}
T^G_{AB}={1\over {4}}g_{AB}F^{MN}F_{MN}-F_A^CF_{BC}\,,
\end{equation}
where Roman letters run over all 5-dimensions. Here, we are only interested in 
determining the contributions to $T^G_{AB}$ made by the lowest gauge KK mode 
in the tower since it is this field that we identify with the usual SM gauge 
boson. 
To do this end we insert the full KK expansion into the above expression and 
concentrate on the lowest mode. Besides the usual 4-d pieces, the field 
strengths $F_{CD}$ in 
the above expressions involve 4-d derivatives of $A_5$ as well as $\phi$ 
derivatives of $A_\mu$. Now we recall that $(i)$ the fifth component of the 
gauge field, $A_5$, is {\it absent} for the zero mode due to the 
orbifold symmetry and $(ii)$ that the zero mode gauge field has a corresponding 
$\phi$-independent bulk wave function. Employing these conditions  
it is easy to see that the new {\it bulk} contributions to 
both $T^{G~\mu}_\mu$ and $T^G_{\mu 5}$ must vanish for these zero-modes. Of 
course it is clear that neither $T^G_{\mu\nu}$ nor $T^G_{55}$ will 
in general vanish. What about these contributions? Again, using $(i)$ and 
$(ii)$ above, all terms proportional to either $A_5$ or $\partial_y A_\mu$ 
can be dropped; the resulting 
action induced by the latter term for the KK zero mode is then given by
\begin{equation}
S^G_{55}=\int d^4x dy \sqrt {-g} ~{{-r}\over {\Lambda_r \pi kr_c}} 
~{-1\over {4}} \eta^{\mu \nu} \eta^{\lambda \tau} e^{4\sigma}F_{\nu \tau}
F_{\mu\lambda}~ {1\over {2\pi r_c}}\,,
\end{equation}
where the fields strengths are now {\it only} those of the zero mode. Here 
we have made use of the notation above and we will further note 
that in the RS model $g=det (g_{AB})=-e^{-8\sigma}$. 
Upon $y$ integration this term in the action leads to a 4-d interaction of 
the generic form 
\begin{equation}
S_4^{(1)}=\int d^4x ~{r\over \Lambda_r} {1\over {4\pi kr_c}} 
F^{\mu\nu}F_{\mu\nu}\,
\end{equation}
with the indices now contracted by the flat space metric; we will return to 
the implications of this term later. 

What about the $T^G_{\mu \nu}$ term? 
Using the definitions above, the corresponding dimension-7 interaction 
term arising from $T^G_{\mu\nu}$ and the 
$\partial_\mu \partial_\nu r$ piece of $g^r_{\mu\nu}$ 
can be symbolically written (after integration over $y$) as 
\begin{equation}
S_4^{(2)}=\int d^4x~
{1\over {\pi^2\sqrt{kr_c}\Lambda_r}} {{(k/\mpl)^{-2}}\over {kr_c\Lambda_\pi^2}}
\sum_n {I_n\over {x_n^2}} \int_\epsilon^1 {{\epsilon^2 dz} \over {z^3}} 
\Big({{z^2 J_2(x_n z)}\over {J_2(x_n)}}-1 \Big)[\partial_\mu \partial_\nu r 
\eta^{\mu\sigma}\eta^{\nu\lambda}T^G_{\sigma\lambda}]\,, 
\end{equation}
where $p^r$ is the radion momentum and we have extracted out only the zero-mode 
contributions to the stress-energy tensor. 
Performing the necessary integrations and sums, this reduces to 4-d interaction 
\begin{equation}
S_4^{(2)}=\int d^4x~ 
{1\over {\Lambda_r}} ~\Big({k\over {0.1\mpl}} \Big)^{-2} ~{0.344\over 
{\Lambda_\pi^2}} [\partial_\mu \partial_\nu \eta^{\mu\sigma}
\eta^{\nu\lambda}T^G_{\sigma\lambda}]\,. 
\end{equation}
Note that we expect $k/\mpl \lsim 0.1$ as discussed above so we have scaled 
the overall numerical factor accordingly. 
Though the above sum extends over an infinite number of terms we have only 
included the first two hundred due to its rapid convergence; 
we have checked that including more terms does not modify our numerical 
results. We will return to the 
phenomenological implications of the actions $S_4^{(1)}+S_4^{(2)}$ below.

\subsection{Bulk Fermions}

Here we are particularly interested in the decay of radions to SM fermions, 
\ie, the three-point function for all particles on-shell. 
Bulk fermions differ from bulk gauge bosons in that a bulk mass term is 
generally present. $SU(2)$ 
doublet (D) and singlet (S) SM fields correspond to different states 
in 5-d with the orbifold symmetry enforcing the correct chirality for the 
corresponding zero modes. The bulk mass terms for these fields have the form 
$sign(\phi)M_D\bar DD+(D\to S)$ where the $sign(\phi)$ factor assures that 
this term in the action is $Z_2$-even. (This factor will be suppressed in the 
discussion that follows.)   
The fermion tower KK expansions can be written quite generally as 
$D=\sum D_L^{(n)}(x)\rho^{(n)}(\phi)+ D_R^{(n)}(x)\tau^{(n)}(\phi)$ and 
$S=\sum S_L^{(n)}(x)\tau^{(n)}(\phi)+ S_R^{(n)}(x)\rho^{(n)}(\phi)$ where the 
$\phi$-dependent $\rho(\tau)$ functions are $Z_2$ even(odd). 
We now see that the bulk mass terms connect the 
left- and right-handed states with the same value of 
$n$ of the form, \eg, $m_n \bar D_L^{(n)} D_R^{(n)} +h.c.$ (and similarly for 
$D\to S$) where $m_n$ is the KK mass. 
(Note that bulk masses {\it do not} connect the $S$ and $D$ towers.)
Since for the $D(S)$ zero mode only a left-(right-)handed state exists, 
there is effectively 
no bulk mass term for this state; hence the bulk mass term will not contribute 
to the stress-energy tensor for zero 
modes. By similar arguments it is also clear that all terms containing a 
$\gamma_5$, the additional gamma matrix in 5-d, must also be absent for zero 
modes since it too connects terms of opposite helicity but with the same 
KK number, $n$. All terms containing 
$\gamma_5$ in the action are of the form 
$\bar D_L^{(n)} D_R^{(n)} +(D\to S)+h.c.$, but only $D_L^{(0)}$ and $S_R^{(0)}$ 
exist for zero modes. Thus terms containing both zero modes and $\gamma_5$ 
are absent. As discussed above, the usual zero mode 
masses are generated on the TeV brane through a coupling of the form 
$\sim \lambda \bar D SH \delta(\phi-\pi)$ via SSB; this wall term 
then provides the `usual' radion coupling to 
fermions in the RS model. As before, we ignore the effects of SSB when 
determining the new bulk contributions to the radion couplings. 
Let us see how these arguments help us to reduce the 
complexity of the potential bulk-induced fermion couplings to radions.

The general form of the stress-energy tensor for a free massive bulk fermion 
field can be written as{\cite {feyn}} 
\begin{equation}
T^f_{AB}=-g_{AB}[\bar \psi i\Gamma^C \partial_C \psi-{1\over 2}
\partial^C (\bar \psi i\Gamma_C \psi)+m \bar \psi \psi]
+{1\over 2}[\bar \psi i\Gamma_A \partial_B \psi -{1\over 2} \partial_B 
(\bar \psi i\Gamma_A \psi)+(A\leftrightarrow B)]\,,
\end{equation}
where $\Gamma_A$ contains the vielbein and $\psi=D,S$. We now inset the full 
KK expansion into the equation above and concentrate only on those terms 
involving the zero mode. The first thing to note is that 
the terms in the first bracket vanish for zero modes for {\it all} 
values of $A,B$. 
This happens for three reasons: $(i)$ when $C=5$, a $\gamma_5$ is present which 
connects $\psi^{(0)}_L$ to $\psi^{(0)}_R$- but only one of these is actually 
present in the theory due 
to the orbifold conditions. $(ii)$ A similar argument applies 
to the contribution from 
the bulk mass term as it connect left- and right-handed chiral modes. 
These two contributions are thus zero. $(iii)$ When $C=\mu$ 
we again get zero by using the bulk equations of motion for the massless 
zero mode: $i\gamma^\mu \partial_\mu \psi=0$. Another way to see the vanishing 
of the first bracket is the use of current conservation for on-shell fields 
and the full 5-d equations of motion. 

The terms in the second 
bracket remain as potential coupling sources; let us first consider forming 
$T^{f ~\mu}_\mu=0$ from it. We see immediately that for  $A,B=\mu,\nu$ the 
second bracket vanishes for zero modes 
when contracted with the 4-d metric due to the bulk equation of motion. 
Thus we see that the bulk contribution yields $T^{f ~\mu}_\mu=0$ for the 
zero modes.  Next we consider $T^f_{55}$; we see however that 
when $A,B=5$ the second bracket  
vanishes due to the presence of the $\gamma_5$ as discussed above. 
There remains only the case $T^f_{\mu 5}$ which is somewhat more subtle. 
In this case $A=\mu,B=5$, two terms immediately 
vanish due the presence of a $\gamma_5$ as before. A third term vanishes by 
using the equations of motion. The last remaining term is then found to be 
proportional to $\sim \bar \psi \gamma_\mu \partial_y \psi$ which doesn't 
immediately vanish. What form of radion interaction can be obtained from this 
term? After integration over 
$y$ and contraction with $g^r_{\mu 5}$, the resulting radion 
interaction is found to be proportional to 
$\sim \eta^{\mu\nu}\partial_\nu r \bar D_L^{(0)}i\gamma_\mu D_L^{(0)}  
+D_L \to S_R$. Due to 4-momentum conservation this vanishes for zero mode  
fermions using the equations of motion as above. 
In a similar fashion one can analyze the dimension-7 interaction term 
$\sim \partial_\mu \partial_\nu r \eta^{\mu\sigma}\eta^{\nu\lambda}
T^f_{\sigma\lambda}$ which can also be shown to vanish using the 
Feynman rules in {\cite {feyn}} for on-shell radions and 
massless zero modes. This removes all
four potential sources for radion on-shell couplings to zero mode 
bulk fermions. This 
implies that the only couplings of radions to fermions arise from the SSB 
Higgs interaction on the TeV 
brane. Thus we conclude that, unlike the case of gauge bosons,  
on-shell zero mode fermions from the RS bulk KK expansion and fermions on the 
TeV brane have identical interactions with radions. The coupling of the radion 
to SM fermions is insensitive to their location.

\section{Phenomenology}

In the last section we obtained the new couplings of the radion to gauge bosons 
in the RS bulk and demonstrated that the corresponding terms for fermions are 
absent. The gauge boson couplings are found to be controlled by the two actions 
$S_4^{(1)}+S_4^{(2)}$ above, in addition to the brane term for generating  
masses of the fermions, as well as $W$ and $Z$ bosons, through the Higgs 
mechanism. The latter brane interaction is the only one present for fermions. 
To begin, it is instructive to examine the form of 
momentum space coupling implied by the $\partial_\mu 
\partial_\nu r\eta^{\mu\sigma}\eta^{\nu\lambda}T^G_{\sigma\lambda}$ 
term in $S_4^{(2)}$. Labelling the momentum space vertex as 
$r(p_r)V(k_1,\epsilon_1^\rho)V(k_2,\epsilon_2^\sigma)$ for generic vector 
bosons $V$ with all momenta flowing into the vertex and using momentum 
conservation we arrive at the following tensor structure for this piece of 
the action
\begin{equation}
-p^r_\mu p^r_\nu \eta^{\mu\sigma}\eta^{\nu\lambda}T^G_{\sigma\lambda}=
\eta_{\rho\sigma}[p_r^2 k_1\cdot k_2-2p^r\cdot k_1 p^r\cdot k_2]+k_{1\sigma}
k_{2\rho}[p_r^2-2k_1\cdot k_2]\,,
\end{equation}
where we have used the Feynman rules in {\cite {feyn}}. Here we have 
dropped terms 
proportional to $k_{1\rho},k_{2\sigma}$ since we are interested in cases 
with on-shell gauge bosons or where one of the gauge bosons is virtual and 
couples to massless fermion pairs. In the case of on-shell 
radion decay to pairs of gauge bosons 
the tensor structure for this vertex simplifies to
\begin{equation}
-M_V^2 m_r^2 \eta_{\rho\sigma}+2M_V^2 k_{1\sigma}k_{2\rho}\,,
\end{equation}
where $M_V$ is the gauge boson mass. (Here we have used the fact that SSB has 
occurred and set $k^2_{1,2}=M_V^2$.) Notice that this part of the decay 
amplitude vanishes when we consider decays to massless gauge bosons. This 
suggests we examine the massive and massless gauge boson radion decays 
separately.

\begin{figure}[htbp]
\centerline{
\includegraphics[width=9cm,angle=90]{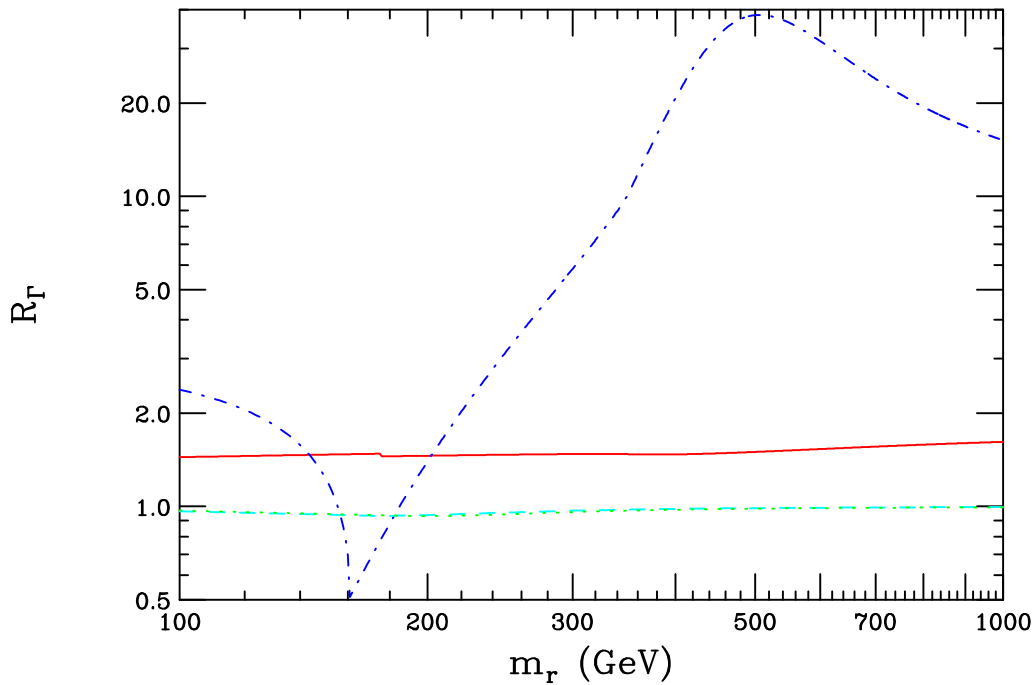}}
\vspace*{0.1cm}
\caption{The ratios for the partial widths of the radion into SM fields 
when they are bulk states 
versus the corresponding case when they are 
wall states as a function of the radion mass. The solid(dashed-dotted, 
dashed,dotted) curve is for the $gg(\gamma\gamma, W^+W^-,2Z)$ final state. 
$\Lambda_r$, whose value only influences the radion decays to massive gauge 
bosons, has been taken to be 1.5 TeV along with $k/\mpl=0.1$. No 
alteration occurs in the case of $\bar ff$ or two Higgs boson final states, 
\ie, $R_\Gamma(\bar ff,hh)=1$.}
\label{fig1}
\end{figure}

\subsection{Massive Gauge Bosons}

Combining the terms that arise from $S_4^{(1)}+S_4^{(2)}$ with the usual wall 
term above{\cite {big}} we obtain the full $rVV$ coupling; the 
matrix element for radion decay to massive gauge 
bosons can be written as 
\begin{equation}
{\cal M}={M_V^2\over \Lambda_r} \epsilon_2^\sigma \epsilon_1^\rho 
[A\eta_{\rho\sigma}+{B\over M_V^2}k_{1\sigma}k_{2\rho}]\,, 
\end{equation}
which leads to the partial decay width 
\begin{equation}
\Gamma_V={M_V^4\over {16\pi~m_r\Lambda_r^2}}(1-4M_V^2/m_r^2)^{1/2}G(x)\,,
\end{equation}
where $x=k_1\cdot k_2/M_V^2=m_r^2/2M_V^2-1$ and the function $G$ is given by 
\begin{equation}
G(x)=A^2(2+x^2)+B^2(1-x^2)^2-2ABx(1-x^2)\,.
\end{equation}
The dimensionless coefficients $A,B$ are given by 
\begin{eqnarray}
A &=& 1-{x\over {\pi kr_c}}-a{m_r^2\over {\Lambda_\pi^2}}\,, \nonumber \\
B &=& {1\over {\pi kr_c}}+2a{M_V^2\over {\Lambda_\pi^2}}\,, 
\end{eqnarray}
with $a=0.344[{0.1\over {k/\mpl}}]^2$ from above. (In our numerical examples 
below we will assume that $k/\mpl=0.1$.) 
We note that in the case of wall gauge fields, $A=1$ and 
$B=0$. From the above expression for $G$ it would at first 
appear that for arbitrary values of $A,B$ this radion partial 
width will grow very rapidly as a high power of $m_r/M_V$ as $m_r$ gets  
large. This is however not the case in reality as these potentially large 
terms cancel with the specific values for $A$ and $B$ that 
we obtained above; this provides a test of our results. 
In fact the growth of this partial width with $m_r$ is found to be 
no greater than that for the case of wall fields.  
Taking the ratio of the above partial width to that obtained in the case of 
gauge fields confined to the TeV brane, $R_\Gamma$, 
we obtain the results in Fig.1. Here we see 
that this ratio is almost flat as a function of $m_r$ with a value near 
$\sim 0.95$ and is found to be quite insensitive 
to $\Lambda_r$. For the entire mass and parameter range 
examined, for both the $W^+W^-$ and $ZZ$ final states, this ratio lies well 
within $\sim 5-10\%$ of unity. 
Clearly these modes alone will not help us test whether or not the SM gauge 
fields lie in the bulk as these partial widths are essentially unaltered.

\subsection{Massless Gauge Bosons}

In this case the only new terms arise from $S_4^{(1)}$ and 
take the same form as those 
obtained from SM loops and the trace anomaly when the SM fields lie on the 
TeV brane{\cite {big}}. These latter terms will still be present as 
will those which arise from loops involving the KK excitations of the SM 
fields. Based on the lower bounds on the KK mass spectrum obtained earlier by 
Hewett, Petriello and Rizzo{\cite {big2}} and the analysis presented in 
{\cite {frank}} by Petriello 
we expect these additional KK loops to give only a very small correction to 
the usual SM contribution and will be subsequently neglected. 

The effective vertex for gluon pairs coupling to the radion is found to be 
\begin{equation}
{1\over {\Lambda_r}}(b_3+{2\over {\alpha_s kr_c}}-{F_g\over {2}}) 
{\alpha_s\over {8\pi}}G_{\mu\nu}G^{\mu\nu}r\,,
\end{equation}
where $b_3=7$ is the $SU(3)$ $\beta$-function, $G^{\mu\nu}$ is the gluon 
field strength 
and $F_g$ is the well-known complex 
kinematic function of the ratio of masses of the  
top quark to the radion found in the case of the analogous Higgs boson 
decay arising from the usual one loop triangle graph. 
Similarly, the radion coupling to two photons is now given by 
\begin{equation}
{1\over {\Lambda_r}}(b_2+b_Y+{2\over {\alpha kr_c}}-F_\gamma) 
{\alpha\over {8\pi}}F_{\mu\nu}F^{\mu\nu}r\,,
\end{equation}
where $b_2=19/6$ and $b_Y=-41/6$ are the 
$SU(2)\times U(1)$ $\beta$-functions and $F_\gamma$ is another well-known 
complex kinematic function of the ratios of the $W$ and top masses to the 
radion mass arising from the one loop triangle graphs. 
In both cases the new terms inversely proportional to $kr_c$ arise from 
the action $S_4^{(1)}$ and can be numerically quite large since they occur 
at the tree level. In the case of gluon pairs, the 
new bulk term increases the $b_3$ contribution by $20-25\%$. In 
the case of photon pairs the new contribution is $\simeq 6$ times larger and 
of the opposite sign than that arising from the beta functions. These new 
contributions may thus lead to drastic changes in the radion partial widths. 
Fig.1 shows the ratio of the partial 
widths for $r\to \gamma\gamma$ and $r\to gg$ for the 
case of bulk gauge fields to the corresponding ones obtained when these 
fields are forced to 
lie on the TeV brane. As expected we see that the width $r\to gg$ receives a 
$40-50\%$ increase over the entire radion mass range. The change in the 
width for $r\to \gamma\gamma$ is much more dramatic, is quite sensitive to 
the value of $m_r$, and can vary by roughly two orders of magnitude over the 
relevant mass range.

\subsection{Collider Signatures}

The results of Fig.1 summarize the differences between radion decays to SM 
fields when they are in the 
bulk or on the TeV brane for various radion partial widths. As far as the 
$\bar ff,~hh$ and, {\it essentially}, 
$W^+W^-$ and $ZZ$ final states are concerned 
there are {\it no} differences. Only the $gg$ and $\gamma \gamma$ modes have 
partial widths which are significantly altered from the TeV brane case. These 
changes in the widths directly led to variations in the various branching 
fractions for decays to bulk fields in comparison to those for wall fields as 
shown in Fig.2. Here we show the deviation of the ratio 
$R_B=B(r\to bulk)/B(r\to brane)$ from unity for the various final states. 
Except for the $\gamma \gamma$ case these ratios of branching 
fractions remain rather flat with increasing $m_r$ above $\simeq 200$ GeV. In 
the case of the $\gamma\gamma$ final state $R_B$ reaches a maximum of 
$\simeq 40$ for radion masses of order 500 GeV. 
Knowing both the width and branching fraction changes for all of the decay 
modes allows one to assess the impact of the `brane vs. bulk' choice for SM 
matter at colliders. For example, for light radions in the 100 GeV or so 
range, the production and decay signature at the LHC would be identical to 
that for a light Higgs, \ie, production by $gg$ fusion followed by decay to 
two photons. From Figs.1 and 2 we see that for radions in this mass range the 
rate for this process is more that twice as large in the case where the SM 
fields are in the bulk than when they are on the TeV brane. On the other 
hand the production of a radion in association with a SM gauge boson followed 
by radion decay to $\bar bb$ would have a rate at the LHC or Linear 
Collider which is smaller in the case of bulk SM states by $\sim 30-35\%$ in 
comparison to the case when the SM is on the wall. If light radions are 
observed at future colliders these small rate differences may help one to 
determine the locations of the various SM fields in the RS model.

\begin{figure}[htbp]
\centerline{
\includegraphics[width=9cm,angle=90]{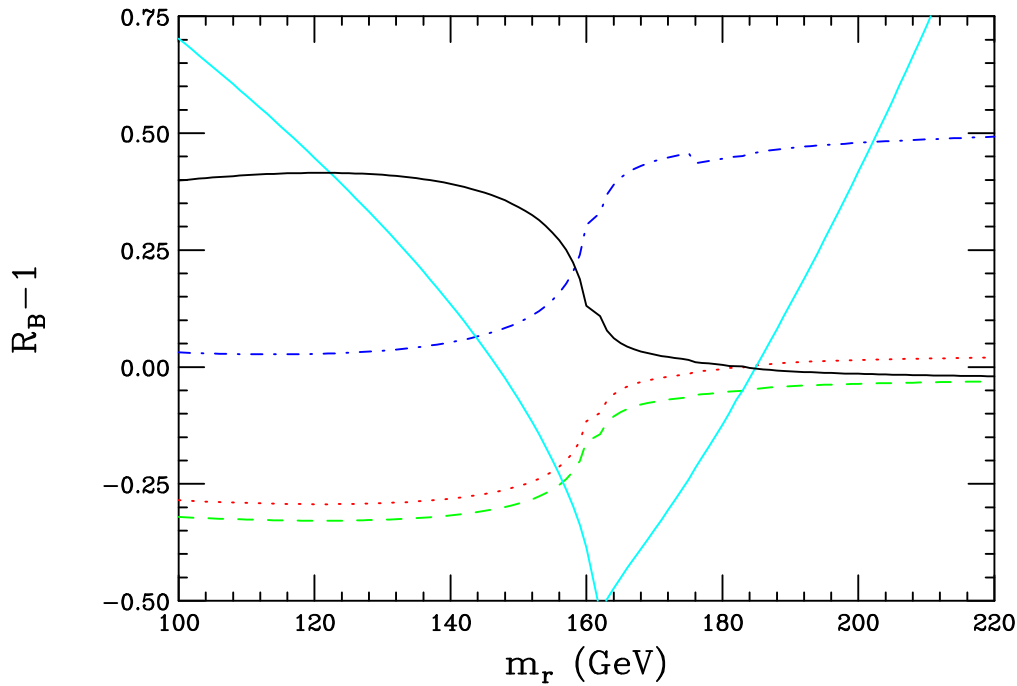}}
\vspace*{0.1cm}
\caption{The ratios for the branching ratios of the radion into bulk states 
versus the corresponding wall states as a function of the radion mass. The 
approximately horizontal curves are, from top to bottom on the left-hand side 
of the figure, for the total width(solid), and the $gg$(dash-dotted), 
$\bar bb$(dots) and $W^+W^-/ZZ$(dashes)  
final states. The V-shaped curve is for the $\gamma\gamma$ final state. 
Radion decay to Higgs boson pairs has been neglected. The values of 
$\Lambda_r$ and $k/\mpl$ are as in the previous figure.}
\label{fig2}
\end{figure}

\subsection{Radion-Higgs Mixing} 

As a last possibility we consider how the mixing of the radion and Higgs 
field may alter the Higgs boson's couplings in the case when the SM fields 
are in the bulk. As is well-known,  
on general grounds of covariance, the radion may mix with the SM Higgs field 
that remains on 
the TeV brane through a wall term in the action of the form 
\begin{equation}
S_{rH}=-\xi \int d^4x \sqrt{-g_w} R^{(4)}[g_w] H^\dagger H\,,
\end{equation}
where $H$ is the Higgs doublet field, 
$R^{(4)}[g_w]$ is the Ricci scalar constructed out of the induced metric $g_w$
on the SM brane,  and $\xi$ is a dimensionless mixing parameter assumed to be 
of order unity and with unknown sign. The 
above action induces kinetic mixing between the `weak eigenstate' $r_0$ and 
$h_0$ fields which can be removed through a set of field redefinitions and 
rotations. This mixing itself is, of course, 
not directly influenced by whether or not the SM fermion 
and gauge fields remain on the TeV brane. 
Clearly, since the radion and Higgs boson couplings to SM fields 
differ and the radion couplings depend on the location of the SM fields 
this mixing will induce modifications in the usual SM expectations for 
the Higgs decay widths and branching fractions which are sensitive to these 
various locations. 

In earlier work by Hewett and Rizzo{\cite {big2}} the influence of 
radion-Higgs mixing on the properties of the Higgs were examined in the case 
where the SM fields were confined to the TeV brane. Those results will be 
somewhat modified if instead the SM fields are now placed in the bulk 
particularly in the case of the $gg$ and $\gamma \gamma$ final states. While 
a detailed study of this phenomenon is beyond the scope of the present 
paper we can get an idea of the size of this effect by examining the shifts 
in the Higgs bosons partial widths in these circumstances. These are shown 
in Fig.3 for a typical 
set of parameter values. Note that these shifts can be significant for values 
of $\xi$ of order unity or less. Comparison with the results of Hewett and 
Rizzo, however, show little qualitative difference between the two possible 
locations of the SM fields. A detailed analysis of these effects will 
be presented elsewhere.

\begin{figure}[htbp]
\centerline{
\includegraphics[width=9cm,angle=90]{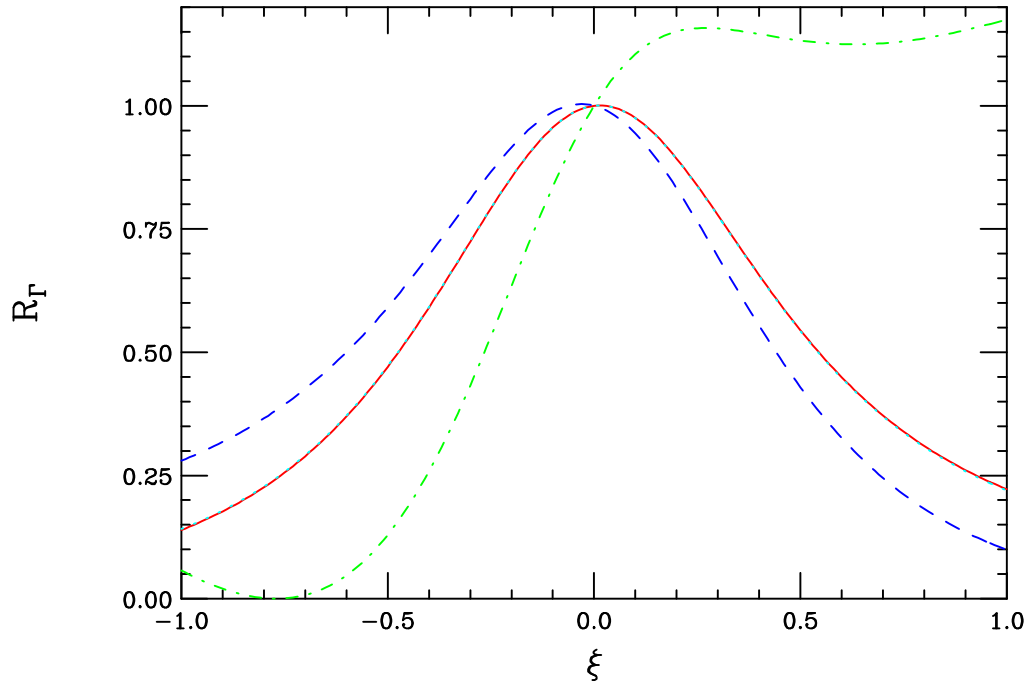}}
\vspace*{0.1cm}
\caption{The effect of mixing on the partial widths of a 125 GeV Higgs 
boson, described by the parameter $\xi$, assuming $v/\Lambda_r=0.2$ and a 
radion mass of 300 GeV as 
discussed in the text. The solid(dash-dotted, dashed, dotted) corresponds to 
the $W^+W^-/2Z(gg, \gamma\gamma, \bar ff)$ final states.}
\label{fig3}
\end{figure}

\section{Conclusions}

In this paper we have examined the couplings of the RS radion to the gauge and 
fermion fields of the SM when these fields lie in the bulk. These results 
were then contrasted to the more conventional case where the SM fields are 
constrained to the TeV brane. Significant differences in decay widths were 
found in the case of the $gg$ and $\gamma\gamma$ final states due to the 
existence of new tree-level terms in the induced 4-d action. Minor differences were 
also found for the $W^+W^-/ZZ$ final states; the corresponding widths for 
decays into fermion or Higgs boson pairs were found to be unchanged. These 
width differences were then shown to lead to relatively large shifts in the 
overall radion branching fractions.

\noindent{\Large\bf Acknowledgements}

The author would like to thank Associate Professor JoAnne L. Hewett and Frank 
Petriello for discussion related to this work.

%
\def\MPL #1 #2 #3 {Mod. Phys. Lett. {\bf#1},\ #2 (#3)}
\def\NPB #1 #2 #3 {Nucl. Phys. {\bf#1},\ #2 (#3)}
\def\PLB #1 #2 #3 {Phys. Lett. {\bf#1},\ #2 (#3)}
\def\PR #1 #2 #3 {Phys. Rep. {\bf#1},\ #2 (#3)}
\def\PRD #1 #2 #3 {Phys. Rev. {\bf#1},\ #2 (#3)}
\def\PRL #1 #2 #3 {Phys. Rev. Lett. {\bf#1},\ #2 (#3)}
\def\RMP #1 #2 #3 {Rev. Mod. Phys. {\bf#1},\ #2 (#3)}
\def\NIM #1 #2 #3 {Nuc. Inst. Meth. {\bf#1},\ #2 (#3)}
\def\ZPC #1 #2 #3 {Z. Phys. {\bf#1},\ #2 (#3)}
\def\EJPC #1 #2 #3 {E. Phys. J. {\bf#1},\ #2 (#3)}
\def\IJMP #1 #2 #3 {Int. J. Mod. Phys. {\bf#1},\ #2 (#3)}
\def\JHEP #1 #2 #3 {J. High En. Phys. {\bf#1},\ #2 (#3)}


\begin{thebibliography}{99}
%
\bibitem{rs} 
L. Randall and R. Sundrum, \PRL 83 3370 1999 .  
%
\bibitem{dhr} 
For an overview of RS phenomenology, see 
H.~Davoudiasl, J.~L.~Hewett and T.~G.~Rizzo, 
Phys.\ Rev.\ D {\bf 63}, 075004 (2001)
[arXiv:hep-ph/0006041]; Phys.\ Lett.\ B {\bf 473}, 43 (2000)
[arXiv:hep-ph/9911262]; Phys.\ Rev.\ Lett.\  {\bf 84}, 2080 (2000)
[arXiv:hep-ph/9909255].
%
\bibitem{gw}
W.D. Goldberger and M. Wise, \PRL 83 4922 1999 ~and \PLB 475 275 2000 ;
C. Csaki, M. Graesser, L. Randall and J. Terning, \PRD D62 045015 2000 ; 
C. Csaki, M. Graesser, and G.D. Kribs, \PRD D63 065002 2001 ; 
C. Charmousis, R. Gregory and V.A. Rubakov, \PRD D62 067505 2000 ;
T. Tanaka and X. Montes, \NPB B582 259 2000 ;
P. Kanti, K.A. Olive and M. Pospelov, hep-ph/0204202. 
%
\bibitem{big}
G.F. Giudice, R. Rattazzi and Wells, \NPB B595 250 2001 ;
U. Mahanta and A. Datta, \PLB B483 196 2000 ;
T. Han, G.D. Kribs and B. McElrath, \PRD D64 076003 2001 ;
M. Chaichian, A. Datta, K. Huitu and Z. Yu, hep-ph/0110035; 
M. Chaichian, K. Huitu, A. Kobakhidze  and Z.-H. Yu, \PLB B515 65 2001 ;
S.B. Bae, P. Ko, H.S. Lee and J. Lee, \PLB B487 299 2000 ; 
S.~Bae, P.~Ko, H.~S.~Lee and J.~Lee, arXiv:hep-ph/0103187;
S.~C.~Park, H.~S.~Song and J.~Song,
Phys.\ Rev.\ D {\bf 65}, 075008 (2002)[arXiv:hep-ph/0103308];
S.~R.~Choudhury, A.~S.~Cornell and G.~C.~Joshi, arXiv:hep-ph/0012043; 
K. Cheung, \PRD D63 056007 2001 ;
U.~Mahanta, Phys.\ Rev.\ D {\bf 63}, 076006 (2001)[arXiv:hep-ph/0008042]; 
J.~L.~Hewett and T.~G.~Rizzo, arXiv:hep-ph/0202155.
%
\bibitem{Kribs}
G.D. Kribs, hep-ph/0110242. 
%
\bibitem{big2}
See, for example, 
A.~Pomarol, Phys.\ Lett.\ B {\bf 486}, 153 (2000)[arXiv:hep-ph/9911294]; 
Y.~Grossman and M.~Neubert, Phys.\ Lett.\ B {\bf 474}, 361 (2000)
[arXiv:hep-ph/9912408]; 
S.~Chang, J.~Hisano, H.~Nakano, N.~Okada and M.~Yamaguchi, 
Phys.\ Rev.\ D {\bf 62}, 084025 (2000)[arXiv:hep-ph/9912498]; 
R.~Kitano, Phys.\ Lett.\ B {\bf 481}, 39 (2000)[arXiv:hep-ph/0002279]; 
S.~J.~Huber and Q.~Shafi, Phys.\ Lett.\ B {\bf 498}, 256 (2001)
[arXiv:hep-ph/0010195] and Phys.\ Rev.\ D {\bf 63}, 045010 (2001)
[arXiv:hep-ph/0005286]; 
S.~J.~Huber, C.~A.~Lee and Q.~Shafi, arXiv:hep-ph/0111465; 
J.~L.~Hewett, F.~J.~Petriello and T.~G.~Rizzo, arXiv:hep-ph/0203091; 
F.~Del Aguila and J.~Santiago, arXiv:hep-ph/0111047, arXiv:hep-ph/0011143, 
Nucl.\ Phys.\ Proc.\ Suppl.\  {\bf 89}, 43 (2000)[arXiv:hep-ph/0011142] and 
Phys.\ Lett.\ B {\bf 493}, 175 (2000)[arXiv:hep-ph/0008143]; 
C.~S.~Kim, J.~D.~Kim and J.~Song, arXiv:hep-ph/0204002.
%
\bibitem{roch}
A.~Das and A.~Mitov, arXiv:hep-th/0203205. See also 
E.~E.~Boos, Y.~A.~Kubyshin, M.~N.~Smolyakov and I.~P.~Volobuev, 
arXiv:hep-th/0202009.
%
\bibitem{feyn}
T.~Han, J.~D.~Lykken and R.~J.~Zhang, Phys.\ Rev.\ D {\bf 59}, 105006 (1999)
[arXiv:hep-ph/9811350]; 
G.~F.~Giudice, R.~Rattazzi and J.~D.~Wells, Nucl.\ Phys.\ B {\bf 544}, 3 (1999)
[arXiv:hep-ph/9811291].
%
\bibitem{frank} 
F.~J.~Petriello, arXiv:hep-ph/0204067 and private communication. 
%
\end{thebibliography}
\end{document}